\documentclass[twocolumn,showpacs,showkeys,amssymb,nobibnotes,superscriptaddress,aps,prl]{revtex4-2}
\usepackage{amssymb}
\usepackage{amsmath}
\usepackage{subeqnarray}
\usepackage{cases}
\usepackage{amsfonts}
\usepackage{bm}
\usepackage{graphicx,braket}
\usepackage[dvipsnames]{xcolor}
\usepackage{calc}
\usepackage{epsfig}
\usepackage{color}
\usepackage{tikz}

\usepackage{amsthm,mathtools}
\usepackage{bbm}

\usepackage[colorlinks,linkcolor=blue,urlcolor=blue,citecolor=blue]{hyperref}

\begin{document}

	\author{Zi-Xu Lu}\thanks{These authors contributed equally to this work}
	\affiliation{Zhejiang Key Laboratory of Micro-Nano Quantum Chips and Quantum Control, and School of Physics, Zhejiang University, Hangzhou 310027, China}
	\author{Gang Liu}\thanks{These authors contributed equally to this work}
	\affiliation{Zhejiang Key Laboratory of Micro-Nano Quantum Chips and Quantum Control, and School of Physics, Zhejiang University, Hangzhou 310027, China}
	\author{Matteo Fadel}
	\affiliation{Department of Physics, ETH Z\"urich, Z\"urich, Switzerland}
	\author{Jie Li}\email{jieli007@zju.edu.cn}
	\affiliation{Zhejiang Key Laboratory of Micro-Nano Quantum Chips and Quantum Control, and School of Physics, Zhejiang University, Hangzhou 310027, China}

\title{Magnonic Gottesman-Kitaev-Preskill states}

\begin{abstract}
		Bosonic quantum error correction encodes a logical qubit in an oscillator, avoiding the hardware overhead of large qubit arrays. Among such encodings, Gottesman-Kitaev-Preskill (GKP) states are paticularly powerful because their phase-space grid structure protects against small displacement errors simultaneously in both conjugate quadratures. Here we provide the first protocol for preparing magnonic GKP states, which involves an ellipsoidal magnetic crystal effectively coupled to a superconducting qubit via a microwave cavity. The geometric anisotropy intrinsically squeezes the magnon mode, while the cavity-mediated qubit control realizes an effective conditional-displacement interaction. We show that two rounds of a conditional-displacement interaction and a qubit projective measurement yield three- and four-component magnonic GKP-like states. We also show how to realize single logical qubit gate operations, such as Pauli, Hadamard and phase gates, completing the logical Pauli basis of the approximate GKP code. 
Our results establish hybrid magnon--qubit systems as a promising platform for preparing bosonic code states, with applications in magnonic fault-tolerant quantum computation and quantum sensing.
	\end{abstract}

\maketitle

	\textit{Introduction.}--The GKP code states~\cite{GKP}, encoding a logical qubit in an infinite-dimensional Hilbert space of a continuous-variable system, represent a powerful approach to quantum error correction~\cite{QEC,Puri}. They are effective in protecting against small errors that occur continually, such as diffusive drifts and amplitude damping of an oscillator or a bosonic mode. Compared to other bosonic codes such as cat~\cite{cat} and binomial codes~\cite{bi}, which are designed primarily to correct excitation loss errors, GKP states exploit displacement operators to protect against shift errors, manifesting as a periodic grid of highly localized peaks in phase space. This structure makes GKP states a vital component of fault-tolerant quantum computation~\cite{FuX,UL21,Dhand21,FuA}, which also find other applications in quantum communication~\cite{Meni20,Loock21,Jiang21,JiangR,Loock24,Jiang25}, quantum sensing~\cite{Wei,Zhuang,Royer26}, and quantum memory~\cite{JiangA,JiangQ}.
	
	Despite their appeal, GKP states are non-Gaussian~\cite{Wal} and notoriously difficult to prepare. Ideal GKP states consist of an infinite superposition of the eigenstates, requiring infinite energy and are thus unphysical; practical implementations rely on finite-component approximations composed of displaced squeezed states~\cite{GKP}. To date, various proposals have been put forward, including approaches based on cat states~\cite{OL}, Fock states~\cite{Ralph}, photon catalysis~\cite{OP}, Kerr interactions~\cite{DV04,Furu22,RoyerA}, generalized Rabi interactions~\cite{ULnpj}, cavity QED~\cite{UL22}, periodic potential~\cite{AL,Di}, shaped free electrons~\cite{IdoX,IdoR}, atomic ensembles~\cite{NC,Vol}, parity measurements~\cite{TB}, fault-tolerant phase estimation~\cite{AC},  neural network~\cite{Nori}, etc. In the experiment, GKP states have only been realized in very limited platforms, including trapped-ion motional modes~\cite{Home19,Home22,Tan24,Tan25}, superconducting microwave cavities~\cite{MD20,MD22,SG22,MD23,Jean}, and propagating optical fields~\cite{Furu}.
	
	Recent years have witnessed a significant development in cavity, nonlinear, and quantum magnonics, due to their broad applications in hybrid quantum systems, quantum information processing, and novel magnonic devices~\cite{Naka19,Li20,Yuan22,Bauer22,qc}. Of particular interest, magnonic systems hold potential applications in quantum sensing~\cite{Tobar19,Crescini, Ruoso20} and quantum computing~\cite{qc}. Nonetheless, quantum computing-related protocols based on magnonics have been rarely explored. In particular, the realization of GKP states in magnonic systems remains to be demonstrated. Here, to bridge this gap, we provide the first protocol to generate magnonic GKP states. The system adopted consists of an ellipsoidal magnetic crystal, whose shape anisotropy intrinsically squeezes the magnon mode, coupled to a superconducting qubit via the mediation of a microwave cavity, which provides nonlinear couplings required for GKP states~\cite{GKP}. Starting from the magnonic squeezed vacuum, two rounds of an effective conditional-displacement (CD) interaction and a qubit projective measurement yield three- or four-component superposition of displaced squeezed states that approximate the square GKP lattice. We first introduce the system, derive the effective CD Hamiltonian, a building block for preparing GKP states, and then show explicitly how magnonic GKP-like states can be generated in our protocol. We further characterize these states by means of Wigner functions, stabilizer expectation values, effective squeezing, and logical fidelity. The CD operation derived in this work allows also for preparing magnonic superposition states and performing magnon characteristic-function tomography~\cite{Home20}.

	\begin{figure}[t]
	\centering
	\hskip-0.5cm\includegraphics[width=0.95\linewidth]{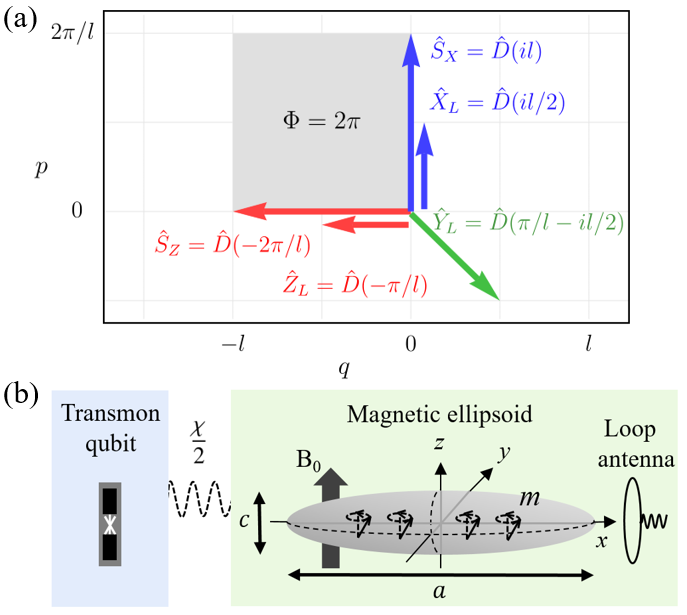}
	\caption{(a) Phase-space representation of the stabilizers $S_X$, $S_Z$ and the logical Pauli operators $X_L$, $Y_L$, and $Z_L$. The enclosed symplectic area sets the commutation relations of the corresponding displacement operators. 
	(b) The squeezed magnon mode in an ellipsoidal magnetic crystal gets effectively coupled to a transmon superconducting qubit via a microwave cavity (not shown). The magnetic crystal is placed inside a uniform bias magnetic field $B_{0}$ along the $z$ axis. A microwave field is loaded via a loop antenna to drive the magnon mode to perform the displacement operation. }
	\label{fig1}
	\end{figure}

		\textit{The GKP code.}--We model magnons, the collective spin excitations in magnetic materials, as a single bosonic mode~\cite{Naka19,Yuan22,Bauer22} with the annihilation (creation) operator $m$ ($m^{\dagger}$), and the associated amplitude and phase quadratures $q=(m^\dagger+m)/\sqrt2$ and $p=(m-m^\dagger)/(i\sqrt2)$, satisfying $[q,p] = i$. We aim to generate magnonic GKP states forming a square lattice in phase space. Since ideal GKP states are non-normalizable, we consider physically realizable approximations constructed from finite superpositions of displaced squeezed states. Specifically, we write the logical basis states as $\ket{0}_L \propto \sum_{k\in\mathbb{Z}}^{k_\text{max}}c_k D(ik \sqrt{2\pi})\ket{r}$ and $\ket{1}_L \propto D(-i\sqrt{\pi/2}) \ket{0}_L$~\cite{GKP,Puri}, where $\ket{r} =  {S}(r)\ket{0}$ is the squeezed vacuum state, $ D(\alpha) = \exp(\alpha  {m}^\dag - \alpha^*  {m})$ is the displacement operator with a complex number $\alpha$, and $S(r)=\exp[(r/2)( m^{\dagger2} - m^2)]$ is the squeezing operator with a real squeezing parameter $r$. The coefficient $c_k$ determines the weight of each displaced component.  Ideal GKP codewords are achieved when $r\rightarrow\infty$ and cutoff $k_\text{max}\rightarrow\infty$~\cite{GKP,Puri}.
	
	The square GKP code encodes a logical qubit in a single oscillator via a lattice structure in phase space defined by commuting displacement operators (Fig.~\ref{fig1}(a)). The stabilizers, {namely the error-check operators,} are chosen as $S_X=D(u)$ and $S_Z=D(v)$, where the primitive lattice vectors $u$ and $v$ satisfy the symplectic condition $\Phi \equiv \mathrm{Im}(v u^*)=2\pi$, ensuring commutativity. Logical Pauli operators correspond to half-lattice displacements $X_L=D(u/2)$, $Z_L=D(v/2)$, and $Y_L=D(-u/2-v/2)$, which anticommute pairwise due to the Weyl relation governing displacement operators. In our protocol, we define the characteristic length $l=\sqrt{2\pi}$, and the effective CD interaction generates displacements along the phase axis, leading to the choice $u=i\,l$ and $v=-2\pi/l$. This construction ensures that logical operators commute with stabilizers while reproducing the Pauli algebra on the encoded subspace, with the resulting code states exhibiting periodicity under the stabilizer lattice translations~\cite{SM}.

	\textit{The system and effective Hamiltonian.}--The system consists of an ellipsoidal magnetic crystal and a transmon superconducting qubit, which are effectively coupled via the mediation of a microwave cavity~\cite{NakamuraSci15,NakamuraSA,NakamuraSci20,Xuda,Xuda2}, as depicted in Fig.~\ref{fig1}(b). The magnetic system is described by the classical magnetic Hamiltonian~\cite{PRL16,Kittel} under a static magnetic field $B_{0}$ along the $z$-axis, which saturates the magnetization $\boldsymbol{M}$ along this direction, i.e., $M_{x,y}\ll M_{z}\approx M_{s}$ ($M_{s}$ is the saturation magnetization), given by
	\begin{equation}
		H=\int_{V_{F}}d^{3}r\left[-\mu_{0}B_{0}M_{z}-\dfrac{\mu_{0}}{2}\boldsymbol{H_{m}}\cdot\boldsymbol{M}+\dfrac{J}{M_{s}^{2}}(\nabla\boldsymbol{M})^{2}\right],
	\end{equation}
	where $V_{F}$ is the volume of the crystal, $\mu_{0}$ is the permeability of vacuum, $J$ is the exchange energy constant, and $\nabla$ is the gradient operator. Here, $\boldsymbol{H_{m}}$ is the demagnetization field due to magnetic dipolar interaction~\cite{de45}: $\boldsymbol{H_{m}}=-(N_{x}M_{x}\boldsymbol{ {x}}+N_{y}M_{y}\boldsymbol{ {y}}+N_{z}M_{z}\boldsymbol{ {z}})$, with $N_{x,y,z}$ being the elements of the demagnetization tensor depending only on the geometric structure. The bias magnetic field $B_{0}$ satisfies $B_{0}+\boldsymbol{H_{m}}\!\cdot \!\boldsymbol{{z}}>0$ to keep the saturation magnetization along the $z$ direction~\cite{de45,B09}.
	
	After quantizing the magnetization via the Holstein–Primakoff transformation~\cite{HP} and performing the Fourier transformation, we obtain the following Hamiltonian for the Kittel mode~\cite{Kittel} in the magnetostatic limit~\cite{SM}:
	\begin{equation}
		H/\hbar=\omega_{m}m^{\dagger}m+\dfrac{\xi}{2}(m^{\dagger2}+ m^{2}),
	\end{equation}
	where $m$ ($m^{\dagger}$) is the annihilation (creation) operator of the Kittel mode with frequency $\omega_{m} =\gamma_0\mu_{0} B_{0}+\left[\gamma_0\mu_{0} M_{s}(1-3N_{z})\right]/2$ and $\xi=\left[\gamma_0\mu_{0}M_{s}(N_{x}-N_{y})\right]/2$ ($\gamma_0$ is the gyromagnetic ratio). 
	For an ellipsoid with $a>b=c$ (Fig.~\ref{fig1}(b)), where $a$, $b$, and $c$ are the ellipsoid semi-axes, the elements of the demagnetization tensor read~\cite{de45}
	\begin{equation}
	N_{x}=\dfrac{1-e^{2}}{2e^{3}}\left(\ln\dfrac{1+e}{1-e}-2e\right),  \,\, N_{y}=N_{z}=\dfrac{1}{2}(1-N_{x}),
	\end{equation}
	with the eccentricity $e=\sqrt{1-(c/a)^{2}}$. This geometric anisotropy leads to a finite parametric term $\xi$, resulting in a magnonic squeezed vacuum state $S(r)|0\rangle$, with $r = \frac{1}{4} \ln[(\omega_{m}+\xi)/(\omega_{m}-\xi)]$. Alternatively, the magnon squeezing can also be achieved by transferring squeezing from external driving fields~\cite{Jie19}, or two-tone driving of a superconducting qubit to induce a two-magnon process~\cite{Guo23,Liu26}.

\begin{figure}[t]
		 \includegraphics[width=0.98\linewidth]{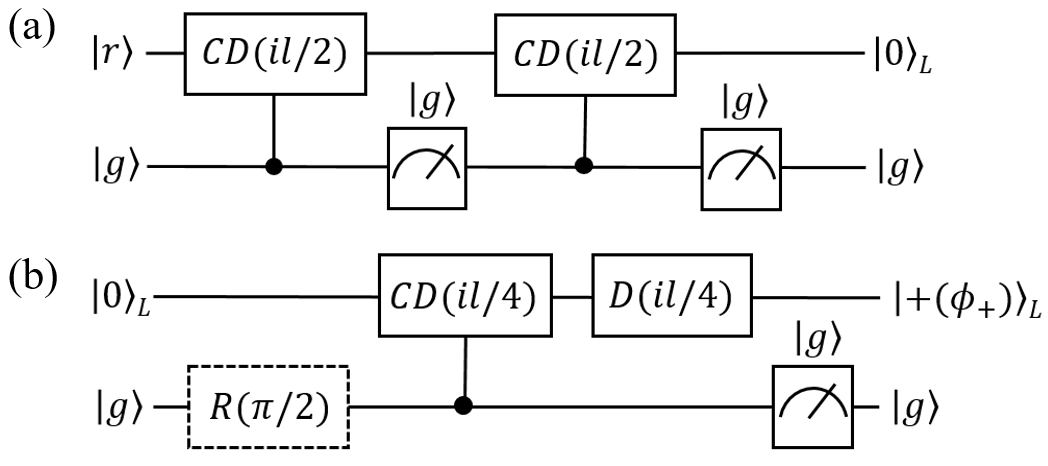}
		\caption{(a) Operation sequence used to generate the GKP state $\ket{0}_L$. The transmon qubit is initially in the ground state $\ket{g}$ and the magnon mode is in the squeezed vacuum state $\ket{r}$. (b) Operation sequence used to perform the logical Hadamard gate (phase gate) onto $\ket{0}_L$ to generate the GKP state $\ket{+}_L$ ($\ket{\phi_{+}}_L$). The ``$CD$" operation represents the magnon-qubit coevolution under the CD interaction, the ``$\ket{g}$" operation denotes the qubit projective measurement, and the ``$D$"  operation stands for the magnon displacement. The qubit $\pi/2$-rotation ``$R$" is only applied for  $\ket{\phi_{+}}_L$.}
		\label{fig2}
	\end{figure}

	The magnon mode is coupled to a transmon qubit via the exchange of virtual photons in a microwave cavity~\cite{NakamuraSci15,NakamuraSA,NakamuraSci20,Xuda,Xuda2}, and the qubit is resonantly driven by a microwave field. By adiabatically eliminating the cavity mode in the large-detuning limit and applying a series of unitary transformations, we obtain the following effective magnon-qubit Hamiltonian which describes a CD interaction~\cite{SM}:
	\begin{equation}
		H_{\mathrm{mq}}/\hbar = -\frac{\chi}{2} (m_s + m_s^\dagger)\sigma_x,
		\label{eq:hamilton}
	\end{equation}
	where $m_s$ ($m_s^\dagger$) is the annihilation (creation) operator of the Bogoliubov mode~\cite{Bo}, whose ground state $\ket{0}_s$ corresponds to a magnonic squeezed vacuum state, and the expression of the coupling strength  $\chi$ is provided in~\cite{SM}. Equation~\eqref{eq:hamilton} indicates a displacement of the magnon mode conditioned on the qubit state: the two eigenstates of $\sigma_x$ with eigenvalues $\pm 1$ correspond to the displacement of the squeezed magnon in opposite directions along the phase axis. 
The displacement amplitude is proportional to $\chi t$, and thus can be controlled by the interaction time.
Combining the CD interaction with a qubit projective measurement, e.g., onto the ground state $\ket{g}$ (Fig.~\ref{fig2}(a)), the magnon mode is projected onto a superposition of two displaced squeezed states. Repeating $N$ rounds of such a displacement-measurement sequence accumulates $2^N$ components in phase space, whose periodic interference gives rise to a comb-like Wigner function~\cite{Shukla2021}---the hallmark of an approximate GKP codeword. 

	\textit{Magnonic GKP-like states.}--We start  with the initial state  $\ket{\psi_0}_s = \ket{0}_s\ket{g}$ (Fig.~\ref{fig2}(a)), where the magnon is in squeezed vacuum $\ket{r} = S(r) \ket{0}_s$ induced by geometry anisotropy, and the qubit is in the ground state $\ket{g}$. 
	The CD Hamiltonian~\eqref{eq:hamilton} corresponds to a unitary evolution operator $U(t_1)$, which, expressed in the $\sigma_x$ eigenbasis $\{\ket{\pm}_x\}$ with $\ket{\pm}_x=(\ket{g} \pm \ket{e})/\sqrt{2}$ and $\sigma_x\ket{\pm}_x = \pm\ket{\pm}_x$, is given by 
	$U(t_1) = \ket{+}_x\!\bra{+}D(i\alpha_{s,1}) + \ket{-}_x\!\bra{-}D(-i\alpha_{s,1})$, where $D(i\alpha_{s,1}) = \exp[i\alpha_{s,1}(m_s + m_s^\dagger)]$ and $\alpha_{s,1} \,{=}\, \chi t_1/2$, with the interaction time $t_1$. The above unitary evolution leads to the joint magnon--qubit state $\ket{\psi(t_1)}_{\rm mq} = [D(i\alpha_{s,1})\ket{0}_s\ket{+}_x + D(-i\alpha_{s,1})\ket{0}_s\ket{-}_x]/\sqrt{2}$, showing explicitly that the qubit is entangled with two oppositely displaced squeezed vacuum wave packets of the magnon mode. 
	Projecting the qubit onto the ground state $ \ket{g}$ leads the magnon mode to a state that is equivalent to applying the operator $E_g(i\alpha_{s,1}) = [D(i\alpha_{s,1}) + D(-i\alpha_{s,1})]/2$ onto the initial state $\ket{0}_s$, i.e., $\ket{\psi}_m = E_g(i\alpha_{s,1})\ket{0}_s$, which is a superposition of two oppositely displaced squeezed vacuum states.

	Repeating the above displacement-measurement operation twice leads the magnon mode to the state $\ket{\psi}_m = E_g(i\alpha_{s,2})E_g(i\alpha_{s,1})\ket{0}_s$, with $\alpha_{s,2} = \chi t_2/2$ and $t_2$ being the duration of the second CD interaction, which can be expanded as a coherent superposition of four components
	\begin{align}\label{eq:psi2expand}
		\ket{\psi}_m 
		=& \frac{1}{4}\bigl\{D[i(\alpha_{s,1}+\alpha_{s,2})] + D[i(\alpha_{s,1}-\alpha_{s,2})] \\
		+& D[-i(\alpha_{s,1}-\alpha_{s,2})] + D[-i(\alpha_{s,1}+\alpha_{s,2})]\bigr\}\ket{0}_s.\nonumber
	\end{align}
	When $t_2 = t_1$, one has $\alpha_{s,2} = \alpha_{s,1} \equiv \alpha_s$, and Eq.~\eqref{eq:psi2expand} reduces to a three-component superposition state $\ket{\psi}_m = \mathcal{N}_3\bigl[D(2i\alpha_s) + 2\mathbbm{1} + D(-2i\alpha_s)\bigr]\ket{0}_s$ with three wave packets centered at $0$ and $\pm 2i\alpha_{s}$. Here the normalization factor $\mathcal{N}_3 = (6 + 8e^{-2|\alpha_s|^2} + 2e^{-8|\alpha_s|^2})^{-1/2}$.
	Alternatively, the choice of $t_2 = 2t_1$ yields $\alpha_{s,2} = 2\alpha_{s,1} \equiv 2\alpha_s$, which produces a four-component superposition state $\ket{\psi}_m = \mathcal{N}_4 [D(3i\alpha_s) + D(i\alpha_s) + D(-i\alpha_s) + D(-3i\alpha_s) ]\ket{0}_s$, with four equally spaced peaks at $\pm i\alpha_s$ and $\pm 3i\alpha_s$, and $\mathcal{N}_4 = (4 + 6e^{-2|\alpha_s|^2} + 4e^{-8|\alpha_s|^2} + 2e^{-18|\alpha_s|^2})^{-1/2}$.

	It should be noted that the CD Hamiltonian~\eqref{eq:hamilton} is derived after a sequence of unitary transformations~\cite{SM}. To examine the magnon state in the laboratory  frame, one has to perform the corresponding  inverse transformations. Specifically, the Fröhlich--Nakajima transformation, used to adiabatically eliminate the microwave cavity, acts as a near-identity operation in the large-detuning limit and its inverse contributes only a negligible correction to the magnon state. The squeezing (Bogoliubov) transformation, when inverted, maps each displaced component back to the laboratory frame but with a rescaled displacement amplitude. The rotating-frame transformation at the drive frequency $\omega_p$ induces a rigid phase-space rotation, which can be removed by choosing the interaction time $t_1$ such that $\omega_p t_1 \simeq 2n \pi$ ($n \in \mathbb{Z}$) and the target displacement condition are simultaneously satisfied.  The remaining transformation acts only on the qubit subspace and has no actual impact on the magnon state.  Taking all these into account, the three-component magnon state takes the following form in the laboratory frame~\cite{SM}
	\begin{equation}
		\ket{\psi'}_m
		= \mathcal{N}_3 \bigl[D(2i\alpha) + 2\openone + D(-2i\alpha)\bigr]\ket{r},
		\label{eq:psi3_lab}
	\end{equation}
	{with $\alpha = \alpha_s e^{-r}$, which represents} a coherent superposition of three equally-spaced squeezed states along the phase axis. 
As for the qubit projective measurements, under the conditions $\omega_p t_1 \simeq 4n\pi$ and $\varepsilon t_1 \simeq 2m\pi$ ($n, m \in \mathbb{Z}$) ($\varepsilon$ is the Rabi frequency)~\cite{SM}, the ground-state projective measurement remains the same in the laboratory frame.  
		The above analysis can be extended straightforwardly  to the four-component case with the choice of $t_2 = 2t_1$, 
		which leads to the magnon state in the laboratory frame given by $\ket{\psi'}_m = \mathcal{N}_4\bigl[D(3i\alpha)+D(i\alpha)+D(-i\alpha)+D(-3i\alpha)\bigr]\ket{r}$.

	\begin{figure}[t]
		\hskip-0.25cm  \includegraphics[width=0.96\linewidth]{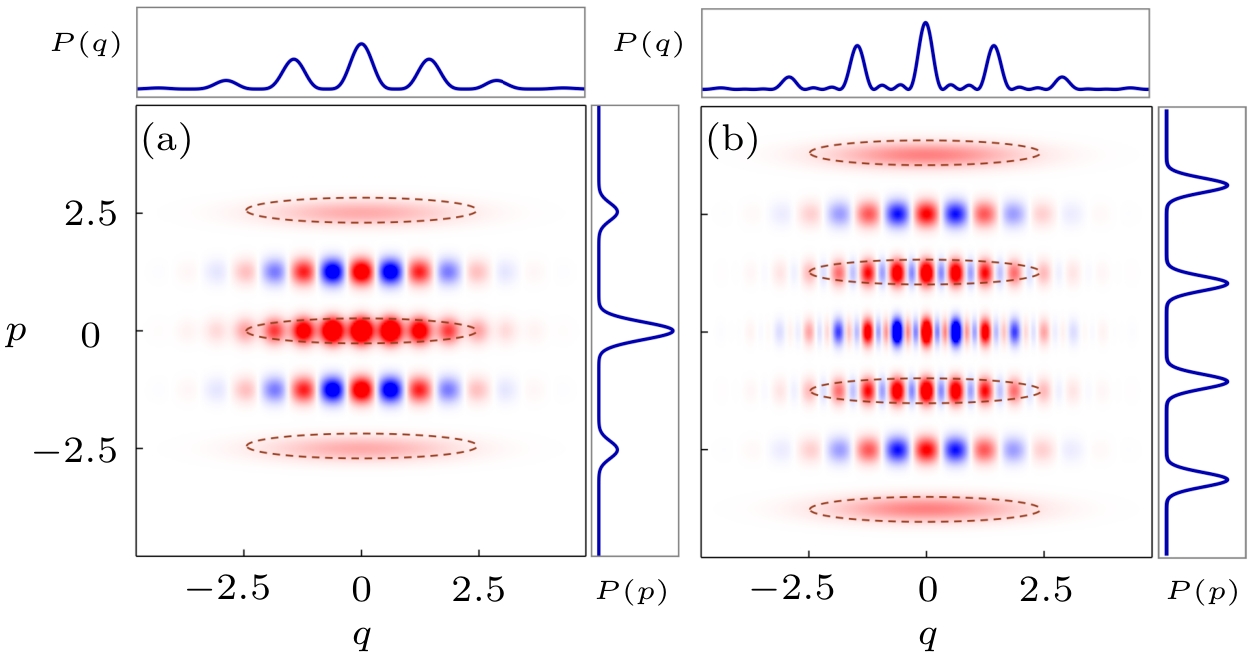}	
	\caption{Magnonic GKP-like states. Wigner function of (a) the three-component superposition state used as the encoded state $\ket{0}_L$, and of (b) the four-component superposition state. Dashed ellipses indicate the centers and characteristic widths of the displaced squeezed components. The side panels show the corresponding marginal Wigner distributions $P(q)$ and $P(p)$. See text for the parameters.}
	\label{fig3}
\end{figure}

\begin{figure}[t]
		\hskip-0.2cm \includegraphics[width=0.95\linewidth]{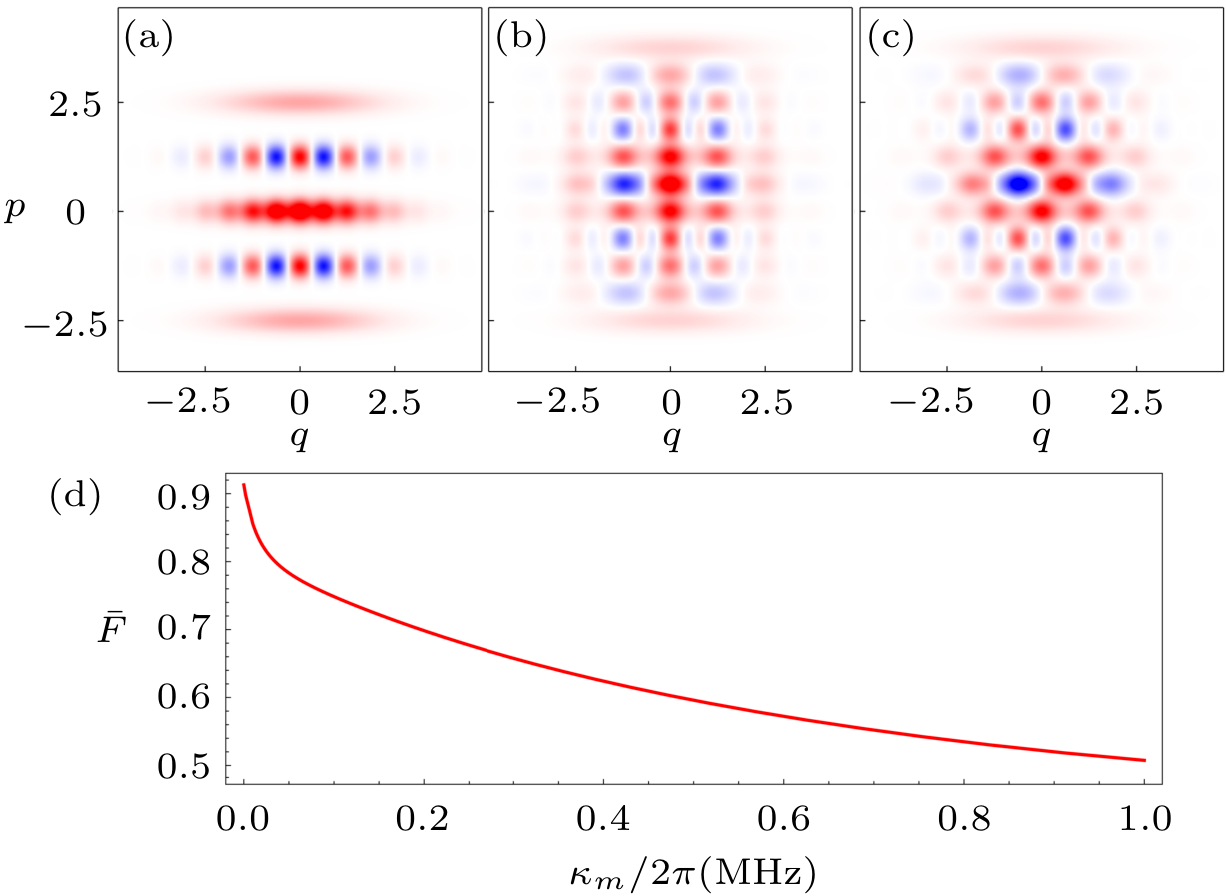}	
		 \caption{Wigner functions of logical Pauli eigenstates: (a) $\ket{0}_L$, (b) $\ket{+}_L$, and (c) $\ket{\phi_{+}}_L$. (d) Average fidelity $\bar{F} $ versus magnon dissipation rate $\kappa_m$.  We consider the three-component superposition state and the parameters are the same as in Fig.~\ref{fig3}(a). We take $\gamma/2\pi = \gamma_\phi/2\pi = 2~\mathrm{kHz}$~\cite{Ren22} and $T = 10~\mathrm{mK}$ in (a)-(d), and $\kappa_m = 5\gamma$~\cite{Chumak25} in (a)-(c). 
		} 
		\label{fig4}
\end{figure}

	Including practical dissipative and dephasing effects, the magnon–qubit system is governed by the following effective master equation~\cite{SM}:
	\begin{align}
		\frac{\partial \rho_\text{mq}}{\partial t}
		\!=& {-} i [ \frac{H_{\mathrm{mq}}}{\hbar},\rho_\text{mq}] +\frac{\kappa_m^\prime}{2}\mathcal{D}[m_s]\rho_\text{mq} +\frac{\kappa_m^{\prime\prime}}{2}\mathcal{D}[m_s^\dagger]\rho_\text{mq}\notag\\
		&+\frac{\gamma^\prime}{2}\mathcal{D}[\sigma_x]\rho_\text{mq}
		+\frac{\gamma^{\prime\prime}}{4}\bigl(\mathcal{D}[\sigma_y]+\mathcal{D}[\sigma_z]\bigr)\rho_\text{mq},
		\label{eq:me31}
	\end{align}
	where $\rho_\text{mq}$ denotes the density operator of the magnon--qubit system, $\kappa_m^\prime = \kappa_m  [(\bar n_{\rm m}+1)\mu^2+\bar n_{\rm m}\nu^2 ]$, $\kappa_m^{\prime\prime} =  {\kappa_m}[(\bar n_{\rm m}+1)\nu^2+\bar n_{\rm m}\mu^2 ]$, $\gamma^\prime = \gamma(2\bar n_q+1)/4$, and $\gamma^{\prime\prime} = [\gamma(2\bar n_q+1)+2\gamma_\phi]/4$, with $\mu = \cosh r$ and $\nu = \sinh r$.~Here, $\kappa_m$ is the magnon damping rate, and $\gamma$ ($\gamma_\phi$) is the qubit dissipation (dephasing) rate.~The mean thermal occupation number $\bar{n}_{m,q} = [\exp(\hbar\omega_{m,q}/k_B T) - 1]^{-1}$, with the bath temperature $T$ and Boltzmann constant $k_B$.~Figures~\ref{fig3} and~\ref{fig4} exhibit our main results of the magnonic GKP-like states without (Fig.~\ref{fig3}) and with (Fig.~\ref{fig4}) the dissipation and dephasing, and the latter is obtained by numerically solving Eq.~\eqref{eq:me31} and applying the inverse transformations to recover the states in the laboratory frame.
	We choose $r \simeq 1$, 
	which yields sufficiently narrow phase-space peaks {with a nearly symmetric extent in} $q$ and $p$~\cite{Home19}. 
	We consider a nearly resonant effective frequency of the qubit and magnon mode about $4.784$~GHz (their original frequencies are provided in~\cite{SM}), and an effective CD coupling strength $\chi/2\pi = 7.55$ MHz. The interaction time for a half-lattice displacement is $t_{1} = t_{l/2} = \sqrt{2\pi}e^{r}/\chi \simeq 143.8$ ns. 

		\textit{Logical encoding.}--The protocol naturally generates a three-component magnon state as a reference encoded state $\ket{0}_L$, satisfying $Z_L\ket{0}_L = +\ket{0}_L$ and $Z_L\ket{1}_L = -\ket{1}_L$. The complementary basis state is obtained by a half-lattice displacement, $\ket{1}_L = X_L\ket{0}_L$, and the remaining four Pauli eigenstates are $\ket{\pm}_L = (\ket{0}_L \pm \ket{1}_L)/\sqrt{2}$ and $\ket{\phi_\pm}_L = (\ket{0}_L \pm i\ket{1}_L)/\sqrt{2}$, where $\ket{\pm}_L$ ($\ket{\phi_{\pm}}_L$) are the eigenstates of $X_L$ ($Y_L$), completing the logical Pauli basis of the approximate magnonic GKP code.  Logical Pauli operations correspond to the magnon displacement which can be implemented by, e.g., applying a resonant microwave drive~\cite{Lu25} via a loop antenna (cf.~Fig.~\ref{fig1}(b)). Logical Hadamard gates can be achieved using the CD interaction, the magnon displacement, and the qubit projective measurement, while phase gates require an additional qubit rotation~\cite{Home19}, as illustrated in Fig.~\ref{fig2}(b).  Figures~\ref{fig4}(a)-\ref{fig4}(c) present the results of the generated logical Pauli eigenstates $\ket{0}_L$, $\ket{+}_L$, and $\ket{\phi_{+}}_L$. The other three eigenstates $\ket{1}_L$, $\ket{-}_L$, and $\ket{\phi_{-}}_L$ can be achieved by performing corresponding displacements~\cite{Home19} and are thus not shown.  

	To quantify the proximity of the generated states to the GKP code space, we evaluate the expectation values of the stabilizers $\langle S_X \rangle$ and $\langle S_Z \rangle$, and extract the corresponding {effective squeezing parameters} $\Delta_j = \sqrt{-\ln|\langle S_j\rangle|^2/(2\pi)}$ ($j\,{=}\,X,Z$)~\cite{Wei}, and the squeezing in decibels is given by $-10\log_{10}(\Delta_j^2)$. Averaging over the six logical Pauli eigenstates yields $\langle S_X\rangle = 0.43$ and $\langle S_Z\rangle = 0.64$, corresponding to the effective squeezing about $5.76~\mathrm{dB}$ and $8.48~\mathrm{dB}$ in the two orthogonal directions. 	  We further characterize the generated states by reconstructing the logical qubit density matrix $\rho_L = (\mathbbm{1} + \langle X_L \rangle X_L + \langle Y_L \rangle Y_L + \langle Z_L \rangle Z_L)/2$, with $\langle \cdot \rangle$ being the expectation value.  The logical qubit quality can be evaluated by calculating the fidelity between the reconstructed state and the corresponding ideal Pauli eigenstate as $F(\rho_L, \ket{\Psi}_L) = \bra{\Psi}_L \rho_L \ket{\Psi}_L$, with $\ket{\Psi}_L$ being one of the six Pauli eigenstates, and we obtain an average fidelity $\bar{F} \simeq 87.3\%$. Note that the finite approximation limits the achievable fidelity up to $91.4\%$.  
Figure~\ref{fig4}(d) indicates that higher fidelity, close to the upper limit  $91.4\%$, can be achieved by further reducing the magnon dissipation rate via, e.g., reducing impurity concentration of the crystal and bath temperature~\cite{Chumak25}.

	In conclusion, we have presented a protocol for magnonic GKP-like states based on a magnon-qubit platform, where the magnon squeezing is induced by geometry anisotropy 
and the CD is implemented by engineering a nonlinear coupling to the qubit.  The work offers the possibility of realizing bosonic quantum error correction on magnonic systems, which finds important applications in magnonic fault-tolerant quantum computation. The magnonic multi-component superposition of squeezed states are useful in quantum sensing, such as the detection of weak magnetic fields and dark-matter axions~\cite{Tobar19,Crescini, Ruoso20}, and the test of unconventional decoherence theories~\cite{Bassi}. The demonstrated CD interaction finds direct applications in preparing other magnonic superposition states, e.g., cat and squeezed Fock states, and performing magnon characteristic-function tomography.

\textit{Acknowledgments.}--JL was supported by Zhejiang Provincial Natural Science Foundation of China (Grant No. LR25A050001), National Natural Science Foundation of China (Grants No. 12474365 and No. 92265202), and National Key Research and Development Program of China (Grants No. 2024YFA1408900 and No. 2022YFA1405200). MF was supported by the Swiss National Science Foundation Ambizione Grant No. 208886, and The Branco Weiss Fellowship -- Society in Science, administered by the ETH Z\"{u}rich.

\setcounter{figure}{0}
\setcounter{equation}{0}
\setcounter{table}{0}
\renewcommand\theequation{S\arabic{equation}}
\renewcommand\thefigure{S\arabic{figure}}
\renewcommand\thetable{S\arabic{table}}

\setcounter{section}{0}

\clearpage    
\onecolumngrid  

\newcommand{\huati}[1]{\mathfrak{#1}}

\section*{Supplemental Materials}

	\section{S1: The Gottesman-Kitaev-Preskill (GKP) code}
	
	Here we provide additional details on the approximate GKP states, the stabilizers, the logical operators, and the associated displacement operators~\cite{GKP}. The displacement operator, defined as
	\begin{equation}
		D(\alpha) = \exp(\alpha {m}^\dagger-\alpha^* {m}),
	\end{equation}
	satisfies the Weyl relation
	\begin{equation}
		D(\beta)D(\alpha)=e^{i\Phi}D(\alpha+\beta),\quad\Phi=\mathrm{Im}(\beta \alpha^*),
	\end{equation}
	which implies the commutation relation
	\begin{equation}
		[D(\alpha),D(\beta)] = 2ie^{i\Phi}\sin(\Phi)D(\alpha)D(\beta).
	\end{equation}
	It indicates that commutation properties are determined by the symplectic area $\Phi$ enclosed in phase space. In particular, two displacements commute when $\Phi=k\pi$ and anticommute when $\Phi=(2k+1)\pi/2$, with $k\in\mathbb Z$. This geometric structure underlies the algebra of the oscillator-based quantum error-correcting codes.
	
	For the square lattice~\cite{GKP,Puri}, we define the characteristic length $l=\sqrt{2\pi}$ and choose lattice vectors $u=il$ and $v=-2\pi/l$, cf. Fig. 1(a) in the main text, yielding
	\begin{equation}
	\begin{aligned}
		&S_X=X_L^{2}=D(il), \\ 
		&S_Z=Z_L^{2}=D(-2\pi/l), \\ 
		&S_Y=Y_L^{2}=D(2\pi/l-il),
		\end{aligned}
	\end{equation}
	which satisfy the required commutation relations: The stabilizers commute with each other ($\Phi=2\pi$) and with logical operators ($\Phi=0,\pi$), while logical operators anticommute pairwise. Although the displacement operators are non-Hermitian in general $ D(\alpha)^\dagger=D(-\alpha)$, their action on the periodic code space ensures that logical operators and their Hermitian conjugates act equivalently, which ensures the correct behavior for Pauli operators.

	\section{S2: Quantization of the Hamiltonian of an ellipsoidal magnetic crystal}
	
	In this section, we show the details on the derivation of the quantized Hamiltonian in Eq. (2) of the main text for an ellipsoidal magnetic crystal. The classical magnetic Hamiltonian in Eq. (1) of the main text is quantized by defining the magnetization operator $\boldsymbol{{M}}=\gamma\boldsymbol{{S}}$~\cite{B09}, where $\gamma$ is the gyromagnetic ratio and $\boldsymbol{{S}}$ is the macrospin density operator. The magnetization is expressed in terms of bosonic excitations via the Holstein-Primakoff transformations~\cite{HP}
	\begin{equation}
		\begin{aligned}
			&M_{+}=\sqrt{2\gamma\hbar M_{s}}[1-\dfrac{\gamma\hbar}{2M_{s}}\tilde{a}^{\dagger}\tilde{a}]\tilde{a},\\
			&M_{-}=\sqrt{2\gamma\hbar M_{s}}\tilde{a}^{\dagger}[1-\dfrac{\gamma\hbar}{2M_{s}}\tilde{a}^{\dagger}\tilde{a}],\\
			&M_{z}=M_{s}-\gamma\hbar\tilde{a}^{\dagger}\tilde{a}.		
		\end{aligned}
	\end{equation} 
	with $M_{\pm}=M_{x}\pm iM_{y}$. Note that the operator $\tilde{a}^{\dagger}\equiv\tilde{a}^{\dagger}(\boldsymbol{r})$~\cite{B09,HP} creates a magnon at position $\boldsymbol{r}$ and satisfies the bosonic commutation relation $[\tilde{a}(\boldsymbol{r}),\tilde{a}^{\dagger}(\boldsymbol{r^{'}})]=\delta(\boldsymbol{r}-\boldsymbol{r'})$. After some calculation, we obtain
	\begin{equation}\label{hhhh}
			H/\hbar=\int_{V_{F}}d^{3}r\bigg\{A\tilde{a}^{\dagger}(\boldsymbol{r})\tilde{a}(\boldsymbol{r})+\left[B^{*}\tilde{a}^{\dagger2}(\boldsymbol{r})+B\tilde{a}^{2}(\boldsymbol{r})\right]+C[\nabla\tilde{a}^{\dagger}(\boldsymbol{r})]\cdot[\nabla\tilde{a}(\boldsymbol{r})]\bigg\},
	\end{equation}
	where
	\begin{equation}
	\begin{aligned}
			&A=\gamma\mu_{0} B_{0}+\dfrac{\gamma\mu_{0} M_{s}(1-3N_{z})}{2},\\ 
			&B=\dfrac{\gamma\mu_{0} M_{s}(N_{x}-N_{y})}{4}, \\
			&C=\dfrac{2\gamma J}{M_{s}}.
	\end{aligned}
	\end{equation}
	The operator $\tilde{a}^{\dagger}(\boldsymbol{r})$ can be expressed in terms of magnon creation operators $m^{\dagger}_{\boldsymbol{k}}$ in the Fourier space  via $\tilde{a}^{\dagger}(\boldsymbol{r})=\sum_{\boldsymbol{k}}\phi^{*}_{\boldsymbol{k}}(\boldsymbol{r})m^{\dagger}_{\boldsymbol{k}}$ with the plane wave eigenstates $\phi_{\boldsymbol{k}}(\boldsymbol{r})=(1/\sqrt{V_{F}})\exp(i\boldsymbol{k}\cdot\boldsymbol{r})$, which leads to  $\nabla\tilde{a}^{\dagger}(\boldsymbol{r})=(-i\boldsymbol{k})\cdot\tilde{a}^{\dagger}(\boldsymbol{r})$ and $\nabla\tilde{a}(\boldsymbol{r})=i\boldsymbol{k}\cdot\tilde{a}(\boldsymbol{r})$. Taking the Fourier transformation of the Hamiltonian~\eqref{hhhh}, we obtain
	\begin{equation}
		H/\hbar=\sum_{\boldsymbol{k}}\sum_{\boldsymbol{k'}}\left[A_{\boldsymbol{k,k'}}m_{\boldsymbol{k}}^{\dagger}m_{\boldsymbol{k'}}+B_{\boldsymbol{k,k'}}^{*}m_{\boldsymbol{k}}^{\dagger}m_{\boldsymbol{k'}}^{\dagger}+B_{\boldsymbol{k,k'}}m_{\boldsymbol{k}}m_{\boldsymbol{k'}}\right],
	\end{equation}
	with 
	\begin{equation}
		\begin{aligned}
			&A_{\boldsymbol{k,k'}}=\dfrac{1}{V_{F}}\int_{V_{F}}d^{3}r\left[\left(A+C\boldsymbol{k}\cdot\boldsymbol{k'}\right)e^{-i(\boldsymbol{k}-\boldsymbol{k'})\cdot\boldsymbol{r}}\right],\\
			&B_{\boldsymbol{k,k'}}=\dfrac{1}{V_{F}}\int_{V_{F}}d^{3}r\left[Be^{i(\boldsymbol{k}+\boldsymbol{k'})\cdot\boldsymbol{r}}\right], \\
		\end{aligned}
	\end{equation}
which after integration gives rise to 
	\begin{equation}
		H/\hbar=\sum_{\boldsymbol{k}}[A_{\boldsymbol{k}}m_{\boldsymbol{k}}^{\dagger}m_{\boldsymbol{k}}+B_{\boldsymbol{k}}^{*}m_{\boldsymbol{k}}^{\dagger}m_{\boldsymbol{-k}}^{\dagger}+B_{\boldsymbol{k}}m_{\boldsymbol{k}}m_{\boldsymbol{-k}}],
	\end{equation}
	with $A_{\boldsymbol{k}}=A_{\boldsymbol{-k}}=A+C\boldsymbol{k}^{2}$ and $B_{\boldsymbol{k}}=B_{\boldsymbol{-k}}=B$. In the magnetostatic limit, we only consider the $\boldsymbol{k}=\boldsymbol{0}$ mode, i.e., the Kittel mode~\cite{Kittel} (defining $m_{\boldsymbol{0}}\equiv m$) with
	\begin{equation}\label{sq}
		H/\hbar=\omega_{m}m^{\dagger}m+\dfrac{\xi}{2}(m^{\dagger2}+ m^{2}),
	\end{equation}
	where
	\begin{equation}
	\begin{aligned}
		\omega_{m}&=\gamma\mu_{0} B_{0}+\dfrac{\gamma\mu_{0} M_{s}}{2}(1-3N_{z}),\\
		\xi&= \dfrac{\gamma\mu_{0}M_{s}}{2}(N_{x}-N_{y}).
	\end{aligned}
	\end{equation}
	
	For a magnetic ellipsoid, its geometric shape is described by $(x/a)^{2}+(y/b)^{2}+(z/c)^{2}=1$ where $a$, $b$, and $c$ are the ellipsoid semi-axes. The corresponding elements of the demagnetization tensor are given by~\cite{de45}
	\begin{equation}
		N_{x,y,z}=\dfrac{1}{2}abc\int_{0}^{\infty}\dfrac{ds}{(j^{2}+s)R_{s}},\ (j=a,b,c)
	\end{equation}
	with $R_{s}=\sqrt{(a^{2}+s)(b^{2}+s)(c^{2}+s)}$ ($s$ is the integral variable). Considering a slender prolate ellipsoid with $a>b=c$, we obtain
	\begin{equation}
	\begin{aligned}
		N_{x} &=\dfrac{1-e^{2}}{2e^{3}}\left(\ln\dfrac{1+e}{1-e}-2e\right),  \\
		N_{y} &=N_{z}=\dfrac{1}{2}(1-N_{x}),
	\end{aligned}
	\end{equation}
	with the eccentricity $e=\sqrt{1-(c/a)^{2}}$.

\begin{figure}[t]
		\centering
		\includegraphics[width=0.5\linewidth]{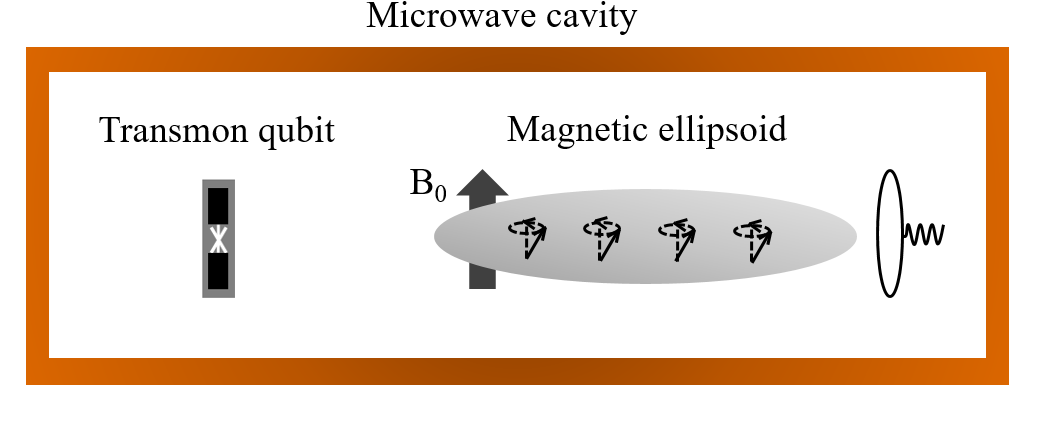}
		\caption{The magnon mode in a magnetic ellipsoid gets effectively coupled to a superconducting qubit via coupling to a common microwave cavity. }
		\label{fig:s1}
	\end{figure}

	\section{S3: Derivation of the effective conditional-displacement Hamiltonian\label{sec3}}

	In this section, we show step by step how the magnon–qubit effective conditional-displacement (CD) interaction is derived.  The effective magnon–qubit  coupling can be established via the mediation of a microwave cavity, as depicted in Fig.~\ref{fig:s1}.  Specifically, the microwave cavity simultaneously couples to the (squeezed) magnon mode in a magnetic ellipsoid and to the superconducting qubit through the magnetic- and electric-dipole interaction, with the coupling strength $g_{cm}$ and $g_{cq}$, respectively~\cite{NakamuraSci15,NakamuraSA}. The Hamiltonian of the cavity–magnon–qubit system is given by  (setting $\hbar=1$) 
	\begin{eqnarray}\label{eq:initial}
		H  =  \frac{\omega_{q}}{2}\sigma_{z} + \omega_{c}c^{\dagger}c + \omega_{m}m^{\dagger}m + \frac{\xi}{2} (m^{\dag2} + m^2)+  g_{cq}(c^{\dagger}+c)\sigma_{x} + g_{cm}(c^{\dagger}+c)(m^{\dagger}+m),
	\end{eqnarray}
	where $c$ ($c^{\dagger}$) and $m$ ($m^{\dagger}$) are the annihilation (creation) operators of the cavity mode and the Kittel mode with frequency $\omega_{c}$ and $\omega_{m}$, respectively, and $\sigma_{z}$, $\sigma_{x}$ are the Pauli operators of the qubit with transition frequency $\omega_{q}$. Note that, here we adopt the Rabi-type coupling form for both the cavity–qubit and cavity–magnon interactions to keep the Hamiltonian in its most general form and to simplify the subsequent derivations following Refs.~\cite{Hwang2015PRL,chen2021NC}.
    By applying a squeezing transformation and defining the magnonic Bogoliubov mode~\cite{Bo}, the Hamiltonian \eqref{eq:initial} becomes in the following form  
	\begin{align}\label{eq:totalS}
		H_1 &= S^\dag(r) H S(r)  \nonumber\\
		&=  \frac{\omega_{q}}{2}\sigma_{z} + \omega_{c}c^{\dagger}c + \omega_{m}^\prime m_s^{\dagger}m_s 
		 +  g_{cq}(c^{\dagger}+c)\sigma_{x} + g_{cm}^\prime(c^{\dagger}+c)(m_s^{\dagger}+m_s)
	\end{align}
	where $S(r) = \exp[\frac{r}{2}(m^{\dag2} - m^2)]$ is the magnon squeezing operator with $r = \frac{1}{4} \ln[1 + {2\xi}/(\omega_{m}-\xi)]$, $\omega_{m}^\prime = (\omega_{m} - \xi) e^{2r}$ and $g_{cm}^\prime = g_{cm} e^{-r}$ are the renormalized frequency and coupling strength associated with the Bogoliubov mode $m_s $. 

	To get an effective coupling between the magnon and the qubit, we adiabatically eliminate the microwave cavity in the large-detuning regime~\cite{NakamuraSci15,NakamuraSA}, which can be realized by taking the Fr{\"o}hlich-Nakajima transformation \cite{Frohlich1950,Nakajima1953}. We write the Hamiltonian~\eqref{eq:totalS} in the form of $H_1 = H_0 + H_I$, with
	\begin{align}
		H_{0} = & \frac{\omega_{q}}{2}\sigma_{z} + \omega_{c}c^{\dagger}c + \omega_m^\prime m_s ^{\dagger} m_s  ,\\
		H_{I} = & g_{cq}(c^{\dagger} + c)(\sigma_{+} + \sigma_{-}) + g_{cm}^\prime(c^{\dagger}+c)( m_s  ^{\dagger} +  m_s ).
	\end{align}
	In the large-detuning regime, where the cavity is far detuned from both the qubit and the magnon mode,
	\begin{align}
		g_{cq}, g_{cm}^\prime \ll |\omega_c-\omega_{q}|, |\omega_c - \omega_{m}^\prime |,
	\end{align}
 the Fr{\"o}hlich-Nakajima transformation takes the form of $H_2 = U_{1}^{\dagger}H_1 U_{1}$, where  $U_1 = e^V$, and $V$ satisfies $H_I+[H_0,V]=0$, given by
	\begin{align}
		V = \mu_q(c\sigma_- - c^\dagger\sigma_+) + \nu_q(c\sigma_+ - c^\dagger\sigma_-)
		+ \mu_m(c   m_s  - c^\dagger  m_s ^\dagger) + \nu_m(c  m_s ^\dagger - c^\dagger  m_s ),
	\end{align}
	with
	\begin{align}
		\mu_q&=\frac{g_{cq}}{\omega_c + \omega_{q}},  \,\,\,\,\,
		\nu_q=\frac{g_{cq}}{\omega_c - \omega_{q}},  \\
		\mu_m&=\frac{g_{cm}'}{\omega_c + \omega_m^\prime },  \,
		\nu_m=\frac{g_{cm}'}{\omega_c - \omega_m^\prime },
	\end{align}
	and we obtain
	\begin{align}
		H_2 = U_{1}^{\dagger}H_1 U_{1} = H_{0} + \frac{1}{2}[H_{I},V] + \frac{1}{3}[[H_{I},V],V] + \cdots.
	\end{align}
	Since the coefficients $\mu_{q,m}$ and $\nu_{q,m}$ are small under the large-detuning condition, which allows us to neglect higher-than-second-order terms of $[H_{I},V]$ in $H_2 $, the rest part remains a good approximation. Assuming the cavity mode is in the vacuum state, which is the case under a low temperature, e.g., 10 mK, 
we obtain the following magnon–qubit  effective Hamiltonian
	\begin{align}\label{eq:Seff}
		H_2  \simeq  \frac{\omega_{q}'}{2}\sigma_{z} + \omega_m^{\prime\prime}  m_s ^{\dagger} m_s -
		\chi( m_s  ^{\dagger}+ m_s  )(\sigma_{+}+\sigma_{-}), 
	\end{align}
	where $\omega_{q}' = \omega_{q} - g_{cq}^{2}[1/\Delta_{cq}^{(-)} - 1/{\Delta_{cq}^{(+)}}]/2, \omega_m^{\prime\prime} = \omega_m^{\prime} - 2\zeta$, $\zeta =  g_{cm}^{\prime2}[1/\Delta_{cm}^{(-)} + 1/\Delta_{cm}^{(+)}]/2$, and $\chi = g_{cq}g_{cm}^\prime[1/\Delta_{cq}^{(-)} +1/ \Delta_{cq}^{(+)} + 1/ \Delta_{cm}^{(-)} +1/ \Delta_{cm}^{(+)}]/2$, with $\Delta_{cm}^{(\pm)} =\omega_{c}\pm\omega_m'$ and $\Delta_{cq}^{(\pm)} = \omega_{c} \pm \omega_{q}$. The third term describes an effective magnon--qubit interaction mediated by the microwave cavity. Note that, in getting~\eqref{eq:Seff} we have neglected an additional small squeezing on the magnon mode due to the presence of the counter-rotating-wave terms in the cavity–magnon coupling (last term of the initial Hamiltonian~\eqref{eq:initial}), which is a very good approximation under current experimental conditions satisfying $\zeta \ll \omega_m''$~\cite{NakamuraSci15,NakamuraSA}.

	By further assuming $\chi \ll \omega_{q}' + \omega_m^{\prime\prime}$, which is typically well satisfied, the Rabi model~\eqref{eq:Seff} reduces to the Jaynes-Cummings (JC) model after the rotating-wave approximation (RWA), given by
	\begin{equation}
		H_{\mathrm{JC}} = \frac{\omega_{q}'}{2}\sigma_z +\omega_m^{\prime\prime} m_s^\dagger m_s - \chi(m_s\sigma_+ + m_s^\dagger \sigma_-).
		\label{eq:SJC}
	\end{equation}
	To construct the CD interaction for creating magnonic GKP-like states, we apply a microwave driving field to the qubit, which is described by the Hamiltonian $H_\text{d} = \varepsilon (\sigma_+ e^{-i\omega_p t} + \sigma_- e^{+i\omega_p t})$, with the drive frequency $\omega_p$ and Rabi frequency $\varepsilon$. In the frame rotating at the drive frequency, i.e., applying a unitary transformation $U_2 = \exp[-i \omega_p( m_s^\dag m_s + \sigma_z/2)t ]$ to $H_\text{JC} + H_\text{d}$, we obtain
	\begin{align}  \label{eq:dJC}
		H_\text{JC}^\prime =&  \frac{\delta_{q}}{2}\sigma_{z} + \delta_{m}m_s^{\dagger}m_s - \chi(m_s\sigma_+ + m_s^{\dagger}\sigma_{-}) + \varepsilon (\sigma_+  + \sigma_- ),
	\end{align}
	with $\delta_{q} = \omega_{q}^\prime - \omega_p$, and $\delta_{m} = \omega_{m}^{\prime\prime} - \omega_p$.  For the resonant case $\delta_{q} = \delta_{m} = 0$, and further in a rotating frame corresponding to the unitary transformation $U_3 = \exp[-i\varepsilon (\sigma_+  + \sigma_- )t]$, the Hamiltonian~\eqref{eq:dJC} becomes  
	\begin{align}  \label{eq:udJC}
		H_{\mathrm{mq}} = - \chi\left\{\left[\frac{\sigma_x}{2} + \frac12(-\sigma_z + i\sigma_y )e^{2i\varepsilon t}  + \frac12(\sigma_z + i\sigma_y )e^{-2i\varepsilon t} \right]m_s + \text{H.c.}\right\}.
	\end{align}
For a relatively strong drive with $2\varepsilon \gg \chi/2$, which is satisfied under our parameters, the fast-oscillating terms in Eq.~\eqref{eq:udJC} can be neglected, which leads to the following effective CD Hamiltonian as used in the main text: 
	\begin{align} \label{eq:Hcd}
	H_{\mathrm{mq}} = -\frac{\chi}{2}(m_s + m_s^\dagger)\sigma_x.
	\end{align}

\section{S4: The GKP states in the laboratory frame}

	As shown in the preceding section, the effective CD Hamiltonian~$H_{\mathrm{mq}}$ is derived through a series of unitary transformations. Therefore, to recover the magnon state in the laboratory frame requires to apply the corresponding inverse transformations.
	The Bogoliubov transformation $S(r)$ introduced to diagonalize the magnon Hamiltonian maps the state in the squeezed-state representation to the original one via $\ket{\psi} = S(r)\ket{\psi}_s$. Specifically, the Bogoliubov ground state $\ket{0}_s$ corresponds to the squeezed vacuum state $\ket{r} = S(r) \ket{0}_s$. 
	The Fr\"ohlich--Nakajima transformation $U_1$ is introduced to adiabatically eliminate the microwave cavity mode in the large-detuning regime, $g_{cq}/\Delta_{cq}^{(-)}, g'_{cm}/\Delta_{cm}^{(-)} \ll 1$, where the cavity remains virtually unexcited. Its impact on the magnon and qubit operators is of orders of $g_{cq}/\Delta_{cq}^{(\pm)}$ and $g'_{cm}/\Delta_{cm}^{(\pm)}$, which is negligible under the large-detuning condition.  Then, $U_1$ acts as a near-identity operation for the magnon--qubit system, and its inverse contributes only a negligible correction to the magnon state.   
	The rotating-frame transformation $U_2$ induces a rigid rotation of the magnon state in phase space at frequency $\omega_p$. By choosing the interaction time such that $\omega_p t = 2n\pi $ ($n \in \mathbb{Z}$),  the coordinate axes of the laboratory frame overlap those of the rotating frame.
	Finally, the transformation $U_3$ acts only on the qubit subspace and thus has no actual impact on the magnon state. Considering only the relevant transformations $S(r)$ and $U_2$, the three-component superposition state $\ket{\psi}_m$ takes the following form in the laboratory frame
	\begin{align}\label{eq:psi3_lab_step1}
		\ket{\psi'}_m
		\simeq S(r)\,U_2 \ket{\psi}_m = \mathcal{N}_3\,S(r)\,U_2
		\bigl[D(2i\alpha_s)+2\openone+D(-2i\alpha_s)\bigr]\ket{0}_s.
	\end{align}
	Utilizing $U_2 D(\alpha_s)U_2^\dagger = D(\alpha_s e^{-i\omega_p t})$, with $t=t_1=t_2$, Eq.~\eqref{eq:psi3_lab_step1} becomes
	\begin{align}\label{eq:psi3_lab_step2}
		\ket{\psi'}_m
		= \mathcal{N}_3\,S(r)\bigl[D(2i\alpha_s e^{-i\omega_p t}) + 2\openone + D(-2i\alpha_s e^{-i\omega_p t})\bigr]\ket{0}_s.
	\end{align}	
	By further using $S(r)D(\alpha_s) S^\dag(r)= D(\alpha_s e^{-r})$, and taking $\omega_p t = 2n\pi $ ($n \in \mathbb{Z}$), 
	we achieve
	\begin{equation}\label{eq:psi3_lab_final}
		\ket{\psi'}_m
		= \mathcal{N}_3\bigl[D(2i\alpha)+2\openone+D(-2i\alpha)\bigr]|r\rangle,
	\end{equation}
	where $\alpha = \alpha_s e^{-r}$ and $|r\rangle = S(r)|0\rangle_s$ is the squeezed vacuum state in the laboratory frame.
 
 For the qubit, both the $U_2$ and $U_3$ transformations imprint a rotating phase. However, under the conditions $\omega_p t = 4n\pi $ and $\varepsilon t = 2 m\pi$ ($m,n\in\mathbb{Z}$), the transformations reduce to the identity on the qubit subspace, so the ground-state projective measurement used in the rotating frame coincides with that in the laboratory frame.


	\section{S5: Derivation of the effective master equation \label{appendix:C}}

	Here we derive the effective master equation corresponding to the effective Hamiltonian $H_{\mathrm{mq}}$. Since $H_{\mathrm{mq}}$ is obtained from the Hamiltonian $H$ in Eq.~\eqref{eq:initial} through a sequence of unitary transformations, the dissipative terms must be transformed accordingly. The master equation of the whole cavity–magnon–qubit system is given by
	\begin{align}
		\dot{\rho} ={}& -i[H,\rho]
		+\frac{\kappa_c(\bar n_c+1)}{2}\mathcal{D}[c]\rho
		+\frac{\kappa_c\bar n_c}{2}\mathcal{D}[c^\dagger]\rho
		+\frac{\kappa_m(\bar n_m+1)}{2}\mathcal{D}[m]\rho
		+\frac{\kappa_m\bar n_m}{2}\mathcal{D}[m^\dagger]\rho \nonumber\\
		&+\frac{\gamma(\bar n_q+1)}{2}\mathcal{D}[\sigma_-]\rho
		+\frac{\gamma\bar n_q}{2}\mathcal{D}[\sigma_+]\rho
		+\frac{\gamma_\phi}{4}\mathcal{D}[\sigma_z]\rho,
		\label{eq:me1}
	\end{align}
	where {$\rho$ is the density operator of the system}, $\mathcal{D}[o]\rho = 2o\rho o^\dagger - \{o^\dagger o,\rho\}$ is the standard Lindblad dissipator, $\kappa_c$ ($\kappa_m$) is the cavity (magnon) damping rate, $\gamma$ ($\gamma_\phi$) is the qubit dissipation (dephasing) rate, and $\bar{n}_{c,m,q} = [\exp(\hbar\omega_{c,m,q}/k_BT)-1]^{-1}$ are the mean thermal occupations, {where $T$ is the bath temperature and $k_B$ is the Boltzmann constant}.

	Applying the same unitary transformations used in Sec.~3 to the dissipators, the Fr{\"o}hlich-Nakajima transformation formally acts also on the magnon and qubit dissipators, but the induced corrections are of orders of $g_{cq}/\Delta_{cq}^{(\pm)}$ and $g'_{cm}/\Delta_{cm}^{(\pm)}$, which, in the large-detuning limit, are negligible. Therefore, the magnon and qubit dissipators are taken as their original forms. 
	The Bogoliubov transformation $S(r)$ reshapes the magnon dissipators and generates an anomalous contribution $G[o]\rho = 2o\rho o - (oo\rho+\rho oo)$. Under the conditions for deriving the Hamiltonian $H_\mathrm{JC}$ in Eq.~\eqref{eq:SJC}, we obtain the corresponding effective master equation, which is
	\begin{align}
		\frac{\partial\rho_\text{mq}}{\partial t}
		={}& -i[H_\mathrm{JC},\rho_\text{mq}]
		+\frac{\kappa_m}{2}\bigl[(\bar{n}_m+1)\mu^2+\bar{n}_m\nu^2\bigr]\mathcal{D}[m_s]\rho_\text{mq}
		+\frac{\kappa_m}{2}\bigl[(\bar{n}_m+1)\nu^2+\bar{n}_m\mu^2\bigr]\mathcal{D}[m_s^\dagger]\rho_\text{mq} \notag\\
		&+\frac{\kappa_m}{2}(2\bar{n}_m+1)\mu\nu\bigl\{G[m_s]\rho_\text{mq}+G[m_s^\dagger]\rho_\text{mq}\bigr\}
		+\frac{\gamma(\bar{n}_q+1)}{2}\mathcal{D}[\sigma_-]\rho_\text{mq}
		+\frac{\gamma\bar{n}_q}{2}\mathcal{D}[\sigma_+]\rho_\text{mq}
		+\frac{\gamma_\phi}{4}\mathcal{D}[\sigma_z]\rho_\text{mq},
		\label{eq:me2}
	\end{align}
	where $\rho_\text{mq}$ is the density operator of the magnon--qubit system, $\mu=\cosh r$, and $\nu=\sinh r$. 
		
	Under the transformation $U_2$, the anomalous terms $G[m_s]\rho_\text{mq}$ and $G[m_s^\dagger]\rho_\text{mq}$ acquire oscillating phase factors $e^{\pm 2i\omega_p t}$, which average to zero under the RWA condition $2\omega_p \gg \{\chi, \kappa_m, \gamma, \gamma_\phi\}$. Similarly, under the transformation $U_3$, the terms oscillating at frequencies $\pm 2\varepsilon$ and $\pm 4\varepsilon$ can be neglected under the strong-driving condition $2\varepsilon \gg \{\chi, \gamma, \gamma_\phi\}$. The resulting effective master equation reads
	\begin{align}
		\frac{\partial \rho_\text{mq}}{\partial t}
		= -i[H_\text{mq},\rho_\text{mq}] + \frac{\kappa_m'}{2} \mathcal{D}[m_s]\rho_\text{mq}
		    + \frac{\kappa_m''}{2}\mathcal{D}[m_s^\dagger]\rho_\text{mq}
		+\frac{\gamma'}{2}\mathcal{D}[\sigma_x]\rho_\text{mq}
		+\frac{\gamma''}{4}\bigl(\mathcal{D}[\sigma_y]+\mathcal{D}[\sigma_z]\bigr)\rho_\text{mq},
		\label{eq:me3}
	\end{align}
	where  we have defined the effective decay rates $\kappa_m' = \kappa_m[(\bar{n}_m+1)\mu^2+\bar{n}_m\nu^2]$, $\kappa_m'' = \kappa_m[(\bar{n}_m+1)\nu^2+\bar{n}_m\mu^2]$, $\gamma' = \gamma(2\bar{n}_q+1)/4$, and $\gamma'' = [\gamma(2\bar{n}_q+1)+2\gamma_\phi]/4$. 

\section{S6: System parameters used for generating GKP-like states   \label{appendix:D}}

In the main text, we provide only the effective frequencies of the qubit and magnon mode and their effective coupling strength $\chi$.  They are however derived quantities and here we provide the corresponding original parameters including also the parameters associated with the cavity. The resonance frequencies of the system are: $\omega_q/2\pi = 4.790$ GHz for the qubit, $\omega_m/2\pi = 18.016$ GHz for the magnon mode, and $\omega_c/2\pi = 5.127$ GHz for the cavity mode. The cavity--qubit and cavity--magnon coupling strengths are $g_{cq}/2\pi = 65~\text{MHz}$ and $g_{cm}/2\pi = 103~\text{MHz}$. The parametric squeezing strength induced by the geometric anisotropy is taken $\xi/2\pi =  17.368~\text{GHz}$, which, together with the magnon frequency $\omega_m$, determines the squeezing parameter $r = \frac{1}{4} \ln[1 + {2\xi}/(\omega_{m}-\xi)] \simeq 1$ as adopted in our protocol. 
The renormalized magnon frequency and cavity--magnon coupling strength are given by
\begin{align}
	\omega_m' &= (\omega_m - \xi) e^{2r} = 2\pi\times 4.788~\text{GHz},\\
	g_{cm}'    &= g_{cm}\,e^{-r}  = 2\pi\times 37.9~\text{MHz}.
\end{align}
The above parameters, after adiabatic elimination of the microwave cavity, give rise to the effective magnon--qubit coupling strength $\chi /2\pi = 7.55$ MHz, using the formula $\chi = g_{cq}g_{cm}^\prime[1/\Delta_{cq}^{(-)} +1/ \Delta_{cq}^{(+)} + 1/ \Delta_{cm}^{(-)} +1/ \Delta_{cm}^{(+)}]/2$, and the effective frequency of the qubit $\omega_q' $ and of the Bogoliubov mode $\omega_m''$, which are assumed to be resonant with the drive frequency $\omega_p$ in our protocol, i.e.,
\begin{align}
	\omega_q' \simeq \omega_m'' \simeq \omega_p = 2\pi\times 4.784~\text{GHz}.
\end{align}
The interaction time ($t=t_1=t_2$) corresponding to the target half-lattice displacement  $\alpha = \chi t/(2e^r) \simeq \sqrt{2\pi}/2$ is $t=143.8$ ns. To further simultaneously (and very well) satisfy
the periodicity conditions $\omega_p t \simeq 4n\pi $ and $\varepsilon t \simeq 2m\pi $ ($m,n\in\mathbb{Z}$), the drive strength should be appropriately chosen and we take the Rabi frequency $\varepsilon/2\pi = 55.63~\text{MHz}$.

\end{document}